\documentclass[
    aps,
    prd,
    superscriptaddress,
    nofootinbib,
    reprint,
    longbibliography,
    floatfix
]{revtex4-2}

\usepackage{amsmath}
\usepackage{amssymb}
\usepackage{mathtools}
\usepackage{empheq}

\usepackage{graphicx}
\usepackage{booktabs}
\usepackage{multirow}
\usepackage{array}

\usepackage[T1]{fontenc}
\usepackage{lmodern}
\usepackage[final]{microtype}
\usepackage{textcomp}

\usepackage[dvipsnames]{xcolor}
\usepackage[autostyle]{csquotes}
\usepackage{orcidlink}
\usepackage[normalem]{ulem}
\usepackage{hyperref}

\hypersetup{
    colorlinks=true,
    citecolor=blue,
    linkcolor=blue,
    urlcolor=blue
}

\newcounter{mnotecount}[section]

\newcommand{\IITGN}{%
    Indian Institute of Technology Gandhinagar,
    Palaj, Gandhinagar 382055, Gujarat, India%
}

\newcommand{\CMI}{%
    Chennai Mathematical Institute, Plot H1 SIPCOT IT Park Siruseri, Chennai 603103, Tamil Nadu, India%
}

\begin{document}

\title{Quasinormal Ringdown and Echoes in Accreting Exotic Compact Objects}

\author{Chanchal Sharma}
\email{chanchal.sharma@iitgn.ac.in}
\affiliation{\IITGN}

\author{Kabir Chakravarti}
\email{kabirc@cmi.ac.in}
\affiliation{\CMI}

\author{Sudipta Sarkar}
 \email{sudiptas@iitgn.ac.in}
\affiliation{\IITGN}

\begin{abstract}
Exotic compact objects (ECOs) may produce late-time gravitational-wave (GW) echoes, but sufficiently compact ECOs undergoing realistic accretion may eventually form an event horizon. We investigate the evolution of quasinormal modes and GW echoes during this transition. Modeling the exterior spacetime with the ingoing Vaidya solution, we numerically evolve scalar perturbations on the resulting dynamical background. We show that the echo signal is progressively compressed and suppressed during accretion and disappears once the growing event horizon engulfs the reflecting surface, after which the waveform smoothly approaches the standard black hole ringdown. Our results identify the disappearance of GW echoes as a characteristic signature of horizon formation and provide a framework for studying dynamical compact objects with future GW observations.\end{abstract}

\maketitle



\section{Introduction}\label{Sec:intro}

Black holes (BHs) are a unique prediction of Einstein's field equations of general relativity (GR). Based on our current understanding, BHs play a central role in a wide range of astrophysical phenomena. Our understanding of BHs, as well as other compact objects such as neutron stars, boson stars, and gravastars, has advanced significantly since the first direct detection of gravitational waves (GWs) by LIGO \cite{LVK:GWTC4}. This is because GWs are among the very few messengers capable of probing the strong-field regime of compact massive objects. With the improved sensitivity expected from future GW detectors, we anticipate observing the oscillations of the merger remnant (not necessarily a BH) with unprecedented precision during the post-merger phase. This so-called ringdown regime can be understood analytically in terms of Quasi-Normal Modes (QNMs) \cite{PhysRev.108.1063,PhysRevLett.24.737, Vishveshwara1970ScatteringOG,PhysRevD.1.2870,1971ApJ...170L.105P,1973ApJ...185..635T,89c00a43-362f-3c19-9cc2-fa0808f91935,337df873-78e6-36e3-92b3-9f9d75e0fe11,Kokkotas_1999}. It is now well established that QNMs consist predominantly of exponentially damped oscillations characterized by a discrete spectrum of complex frequencies. Consequently, considerable attention has recently been devoted to understanding QNMs as probes of gravity in different spacetimes and within modified theories of gravity. Unsurprisingly, QNMs probe not only the underlying gravitational theory and the surrounding geometry but also the nature of the oscillating compact object itself. In particular, they can potentially reveal whether the object possesses an event horizon or a physical surface through the presence or absence of post-ringdown oscillatory features known as GW echoes \cite{2017NatAs...1..586C}. Compact objects with a horizonless surface constitute a subclass of the broader family of Exotic Compact Objects (ECOs). Although many studies have identified various theoretical pathologies associated with horizonless ECOs, they have not yet been ruled out conclusively. As a result, ECOs remain viable theoretical constructs and continue to be subject to both theoretical and observational scrutiny. As we demonstrate later, such GW echoes play a central role in the analysis presented in this work.

Despite this significant progress, most theoretical studies of QNMs and GW echoes have been developed on one of the relatively few exact background solutions of Einstein's field equations. In almost all such studies, the background spacetime is assumed to be either static or stationary. Astrophysical compact objects, however, are often dynamical, particularly during episodes of accretion or mass loss. In this context, the Vaidya solution is especially important, as it is the only known exact spherically symmetric solution describing a fully dynamical spacetime sourced by null radiation \cite{Vaidya:1951fdr}. The Vaidya metric provides a remarkable description of a spacetime undergoing the accretion of matter and/or radiation onto a black hole or compact object. Since accretion generally increases the mass and, more generally, the angular momentum of a compact object, the various characteristic surfaces associated with it evolve with time. Understanding the evolution of these surfaces is particularly interesting for horizonless compact objects. This scenario was investigated by the authors of \cite{Carballo-Rubio:2018vin}, who showed that, for astrophysically relevant accretion rates in an ECO, whose exterior geometry is described by the Vaidya metric, the apparent horizon can overtake the ECO surface within a finite time. The QNMs of the Vaidya black hole spacetime have been studied in the work \cite{Abdalla:2006vb} with the aim of distinguishing them from those of the Schwarzschild spacetime, assuming accretion through radially infalling null radiation.

In this work, we build upon the ideas developed in these two studies \cite{Abdalla:2006vb, Carballo-Rubio:2018vin}. We consider an initially horizonless exotic compact object (ECO) whose exterior geometry is described by the ingoing Vaidya solution, representing continuous accretion of null radiation. Throughout this work, we refer to such a dynamical configuration as a Vaidya ECO. Our central question is the following: how does the event horizon generated during accretion evolve relative to the physical surface of the ECO, and what distinctive signatures does this evolution imprint on the quasinormal-mode (QNM) spectrum and the associated gravitational-wave (GW) echoes? Since late-time GW echoes originate from repeated reflections of perturbations between the effective potential barrier near the photon sphere and the ECO surface, their properties depend sensitively on the evolving separation between these two structures. As accretion proceeds and the horizon approaches and eventually overtakes the surface, this cavity is continuously modified, leading to a characteristic evolution of the late-time echo signal. Our primary objective is therefore to investigate the QNM and echo phenomenology of such Vaidya ECOs throughout the horizon-formation process. This provides a dynamical picture of how the ringdown response evolves during the transition from a horizonless compact object to a black hole, while also exploring the possibility of probing this process through future gravitational-wave observations.

The rest of the paper is organized as follows. In Section \ref{Sec:rev_vg}, we review the key features of the Vaidya geometry. These are then specialized to the case of accreting ECOs in Section \ref{Sec:Acc_Eco}. In Section \ref{Sec:Num_Impl}, we describe the numerical implementation in light-cone coordinates used to compute the QNMs and GW echoes. Our results are presented in Section \ref{Sec:Res}, and we conclude with a discussion in Section \ref{Sec:Disc}.

\textit{Notational Convention} In this work, we use geometric units given by $G=c=1$. For the metric we use the maximal positive sign convention Diag$(-1,+1,+1,+1)$.

\section{Review of the Vaidya Geometry}\label{Sec:rev_vg}
In this section, we review the fundamental geometric frameworks describing stationary and dynamical spherically symmetric black holes and outline the double-null formalism used to study wave propagation and quasinormal ringing in these backgrounds. In General Relativity, the unique spherically symmetric vacuum solution is the Schwarzschild spacetime. In the standard Schwarzschild (or Boyer-Lindquist-type) coordinates $\left(t,r,\theta,\phi\right)$, the metric possesses a well-known coordinate singularity at the event horizon $r=2M$. Although the spacetime manifold itself remains perfectly regular at this hypersurface, the metric components become singular in these coordinates. To remove this coordinate artifact and construct a coordinate system that remains regular across the horizon, it is standard to introduce ingoing Eddington-Finkelstein coordinates, in which the Schwarzschild line element takes the form
\begin{equation}
    \mathrm{d}s^2 = -\left(1 - \frac{2M}{r}\right)\mathrm{d}v^2 + 2 \mathrm{d}v \mathrm{d}r + r^2 \mathrm{d}\Omega^2~,
\end{equation}
where $\mathrm{d}\Omega^2=\mathrm{d}\theta^2+\sin^2\theta\,\mathrm{d}\phi^2$ is the metric on the unit two-sphere, and $v=t+r^*$ denotes the advanced time coordinate, with $r^*$ being the tortoise coordinate. In these coordinates, ingoing radial null geodesics, representing massless particles or inward-propagating radiation, are simply characterized by constant values of $v$ \cite{Poisson:2004}.

Building on this static background, it is of considerable physical interest to study time-dependent spacetimes that describe realistic astrophysical situations in which a black hole changes its mass through accretion or emission of matter and radiation. Such non-stationary processes are naturally described by the Vaidya spacetime. The Vaidya metric generalizes the Schwarzschild solution by promoting the constant mass parameter $M$ to a time-dependent mass function $m(v)$, thereby describing a spherically symmetric geometry sourced by null radiation. For an accreting black hole absorbing an ingoing radial flux of unpolarized null fluid, the mass function increases monotonically, satisfying $dm/dv>0$. The corresponding line element is
\begin{equation}
    \mathrm{d}s^2 = -\left(1 - \frac{2m(v)}{r}\right)\mathrm{d}v^2 + 2\mathrm{d}v\mathrm{d}r + r^2\mathrm{d}\Omega^2~.
\end{equation}

For the analysis of time-dependent wave propagation and quasinormal ringing, however, it is considerably more convenient to reformulate the geometry in double-null coordinates. Unfortunately, an explicit transformation from Eddington-Finkelstein to double-null coordinates is not known analytically for generic mass functions $m(v)$. To circumvent this difficulty, the authors of Refs.~\cite{PhysRevD.34.2978,PhysRevD.70.084014,Abdalla:2006vb} proposed a semi-analytical construction in which the Vaidya spacetime is formulated directly in double-null coordinates \emph{ab initio}, thereby avoiding the need for an explicit coordinate transformation. Within this framework, the spherically symmetric line element is written as
\begin{equation}
    \mathrm{d}s^2 = -2 f(u,v)\mathrm{d}u\mathrm{d}v + r^2(u,v)\mathrm{d}\Omega^2~,
\end{equation}
where $f(u,v)$ and $r(u,v)$ are smooth, non-vanishing functions of the retarded and advanced null coordinates $u$ and $v$, respectively.

The matter source driving the dynamical evolution is modeled, within the eikonal approximation, as a unidirectional flux of unpolarized null radiation with energy-momentum tensor
\begin{equation}
    T_{ab}=\frac{1}{8\pi}h(u,v)k_a k_b~,
\end{equation}
where $k_a$ is the radial null vector. For an ingoing null fluid propagating along the $v$-direction, the Einstein field equations reduce to the coupled system
\begin{equation}\label{eq:for_f}
    f(u,v)=2B(v)\partial_u r(u,v)~,
\end{equation}
\begin{equation}\label{eq:for_r}
    \partial_v r(u,v)=-B(v)\left(1-\frac{2m(v)}{r(u,v)}\right)~,
\end{equation}
\begin{equation}\label{eq:for_h}
    h(u,v)=-4\frac{B(v)m'(v)}{r^2(u,v)}~,
\end{equation}
where a prime denotes differentiation with respect to the advanced time $v$. The functions $m(v)$ and $B(v)$ remain arbitrary, subject only to the weak energy condition,
\begin{equation}
    B(v)m'(v)\le0~.
\end{equation}
For a genuinely dynamical spacetime with $m'(v)\neq0$, one may choose
\begin{equation}
    B(v)=-\frac{1}{2}\,\mathrm{sign}(m')~,
\end{equation}
under which the function $m(v)$ acquires the natural interpretation of the black hole mass, while the coordinate $v$ coincides with the proper time measured by an observer at asymptotic null infinity. Furthermore, the weak energy condition guarantees that $m(v)$ evolves monotonically, ensuring that the null radiation remains consistently ingoing or outgoing throughout the evolution.

The solution of Eqs.~(\ref{eq:for_f})--(\ref{eq:for_h}) therefore determines the Vaidya geometry in double-null coordinates. In practice, the semi-analytical approach reduces to numerically constructing the functions $r(u,v)$, $f(u,v)$, and $h(u,v)$ from these equations. Along each surface of constant $u$, Eq.~(\ref{eq:for_r}) constitutes a first-order ordinary differential equation in $v$. Consequently, the areal radius $r(u,v)$ can be obtained throughout the computational domain by integrating this initial-value problem from a prescribed reference hypersurface $r(u,v_0)$, thereby establishing the causal background on which the perturbations evolve.

The specification of this initial null hypersurface is, in general, a subtle aspect of the construction. For constant $m$ and $B(v)=-1/2$, Eq.~(\ref{eq:for_r}) can be integrated analytically to yield
\begin{equation}\label{eq:schwarztortoisecoord}
    r(u,v)+2m\ln\left|\frac{r}{2m}-1\right|-\frac{v}{2}=P(u)~,
\end{equation}
where $P(u)$ is an arbitrary function of $u$. Throughout the present work, we choose $P(u)=-u/2$, and Eq.~(\ref{eq:schwarztortoisecoord}) provides the initial condition for $r(u,v_0)$. Once $r(u,v)$ has been determined, the remaining functions $f(u,v)$ and $h(u,v)$ follow directly from Eqs.~(\ref{eq:for_f}) and (\ref{eq:for_h}), respectively, for any prescribed mass function $m(v)$.

Having constructed the dynamical background spacetime using this semi-analytical procedure, the evolution of test fields can be investigated by considering linear perturbations propagating on the resulting geometry. In the double-null formulation, scalar perturbations satisfy the wave equation
\begin{equation}
    \frac{\partial^2\psi}{\partial u\partial v}+V(u,v)f(u,v)\psi=0~,
\end{equation}
where the effective potential is given by
\begin{equation}
    V(u,v)=\frac{\ell(\ell+1)}{2r^2(u,v)}+\frac{m(v)}{r^3(u,v)}~.
\end{equation}

The resulting characteristic initial-value problem is solved numerically using the second-order characteristic integration scheme developed by Gundlach, Price, and Pullin \cite{PhysRevD.49.883}. The evolution is performed on a double-null grid with initial data specified on two intersecting null hypersurfaces. Since the late-time quasinormal response is largely insensitive to the precise form of the initial perturbation, we prescribe Gaussian initial data on one null hypersurface,
\begin{equation}
    \psi(u_0,v)=\exp\left[-\frac{(v-v_c)^2}{2\sigma^2}\right]~,
\end{equation}
while a constant initial condition is imposed on the complementary null hypersurface, $\psi(u,v_0)=\mathrm{constant}$, which is taken to be zero in all our numerical calculations. The above formulation forms the basis of all subsequent numerical analyses presented in this work. In the following section, we introduce the specific spacetime model employed in our quasinormal-mode analysis.

\section{Accreting ECO geometry}\label{Sec:Acc_Eco} 

In this section, we describe the dynamical spacetime model adopted to investigate the evolution of quasinormal modes during the transition of an initially horizonless compact object into a black hole. While the causal formation of horizons in horizonless compact objects undergoing accretion has been investigated in previous studies \cite{Carballo-Rubio:2018vin}, the corresponding evolution of the spacetime's quasinormal response has received comparatively little attention. Since quasinormal modes encode the characteristic oscillatory response of a perturbed spacetime and are determined by the underlying geometry, they provide a natural and sensitive probe of the dynamical emergence of a black hole from an initially horizonless configuration. Motivated by this, we investigate how the quasinormal ringing evolves throughout the entire horizon-formation process and examine the signatures associated with the transition from a compact object to a black hole.

We consider an initially horizonless ECO undergoing continuous accretion. As matter and radiation are absorbed by this compact object, its physical surface evolves in response to the incoming energy flux. In Ref.~\cite{Carballo-Rubio:2018vin}, it was shown that sufficiently compact horizonless objects cannot, in general, preserve their horizonless nature under sustained accretion. Remarkably, this conclusion follows purely from causality and is therefore largely independent of the internal structure of the compact object. In particular, no assumptions regarding the equation of state, the composition of the object, the detailed microphysics governing its surface, or even the symmetries of the spacetime are required. The only essential requirement is that the evolution of the physical surface remains causal, namely that it follows a timelike trajectory, with the limiting null evolution corresponding to the maximum physically admissible rate of expansion. This causal restriction determines the largest spacetime region that can respond to the incoming accretion flow. As demonstrated in Ref.~\cite{Carballo-Rubio:2018vin}, sufficiently large accretion rates inevitably exceed this causally allowed expansion, leading to the formation of a trapping horizon. Consequently, the more compact the initial configuration, the smaller the critical accretion rate required to trigger horizon formation.

To investigate the quasinormal-mode evolution associated with this horizon-formation scenario, we consider a spherically symmetric spacetime described by the Vaidya solution reviewed in the previous section. Initially, the ECO possesses a physical surface located outside the trapping horizon, satisfying $r_{ECO}(v)>2m(v)$, where $r_{ECO}(v)$ denotes the radius of the ECO surface. The exterior spacetime is described by the Vaidya geometry, while the interior is left unspecified, since our analysis depends only on the evolution of the surface and the exterior geometry. As accretion proceeds, the mass of the spacetime increases, causing the trapping horizon to expand. Once the trapping horizon overtakes the timelike surface of the ECO, the spacetime undergoes a transition from a horizonless ECO to a black hole.

\begin{figure}[h]
    \includegraphics[width=0.92\columnwidth]{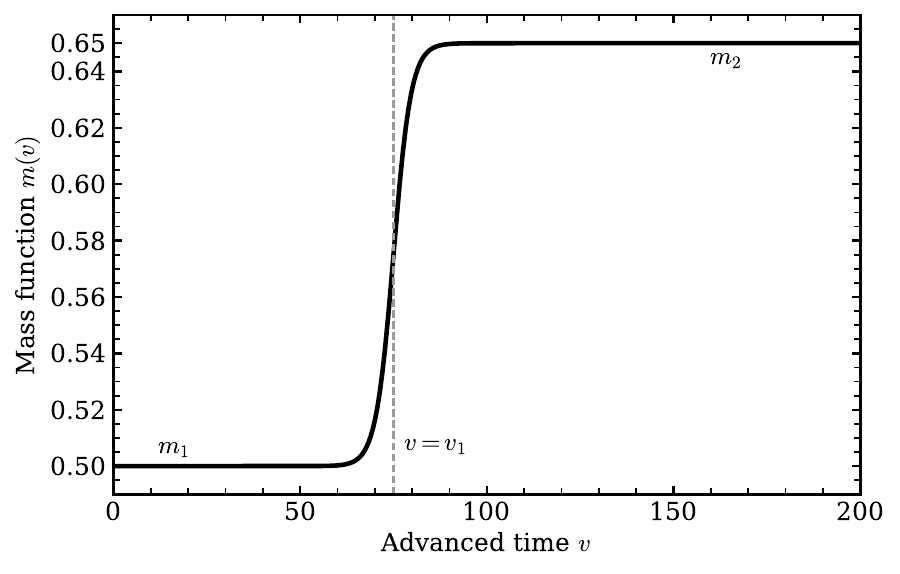}
    \caption{Accretion profile with $m_1=0.5$, $m_2=0.65$, cf. Eq.~(\ref{eq:massprofile}). The dashed vertical line indicates the epoch $v=v_1$.}
    \label{fig:mass_profile}
\end{figure}

The accretion process is modeled through the smooth mass profile
\begin{equation}\label{eq:massprofile}
    m(v)=m_1+\frac{m_2-m_1}{2}\left[1+\tanh\left(\rho(v-v_1)\right)\right]~,
\end{equation}
where $m_1$ and $m_2$ denote the initial and final masses of the compact object, respectively. The parameter $v_1$ specifies the epoch at which the accretion rate reaches its maximum, while $\rho$ controls the duration of the accretion process. The corresponding mass evolution is shown in Fig.~\ref{fig:mass_profile}. The hyperbolic tangent profile provides a smooth interpolation between the initial and final stationary configurations, thereby avoiding discontinuities in the geometry and offering a physically reasonable description of the accretion-driven transition from an initially horizonless ECO to a final Schwarzschild black hole.

The response of the ECO to the accreting matter is modeled by prescribing a timelike trajectory
\begin{equation}\label{eq:ECOsurfaceevol}
    \frac{\mathrm{d}r_{ECO}}{\mathrm{d}v}
    =
    \frac{1}{2}\left(1-\frac{2m(v)}{r_{ECO}}\right)-\epsilon~,
\end{equation}
where $\epsilon$ is a small positive constant that quantifies the deviation from the limiting null evolution. The trajectory of the compact object surface is obtained by integrating this equation subject to an appropriate initial condition specifying the initial surface radius. It is worth noting that, for constant accretion-rate models, causality considerations imply that the formation of a trapping horizon can be avoided only if the compactness parameter satisfies \cite{Carballo-Rubio:2018vin}
\begin{equation}
    \mu \equiv \frac{r_{ECO}-2M}{r_{ECO}}\geq 4\dot{m},
\end{equation}
where $\dot m$ denotes the accretion rate measured by asymptotic observers. In the present work, however, the accretion profile is explicitly time dependent, so that the formation of a trapping horizon cannot be characterized by a single constant parameter and must instead be determined from the full dynamical evolution. Nevertheless, the relevant quantity governing the evolution remains the maximum instantaneous accretion rate,
\begin{equation}
    \dot m(v)=\frac{m_2-m_1}{2}\rho\,\mathrm{sech}^2[\rho(v-v_1)],
\end{equation}
which attains its maximum value
\begin{equation}
    \dot m_{\rm max}=\frac{m_2-m_1}{2}\rho
\end{equation}
at $v=v_1$. For the parameters adopted in the present work,
\[
m_1=0.5,\qquad
m_2=0.65,\qquad
v_1=75,\qquad
\rho=0.2112,
\]
one finds
\begin{equation}
    \dot m_{\rm max}=1.584\times10^{-2},
\end{equation}
which is sufficiently large that the compact object surface eventually becomes enclosed by the growing horizon in our numerical evolution.

Although the causal argument outlined above is formulated in terms of the formation of a trapping horizon, the analysis of quasinormal modes requires knowledge of the event horizon, since it determines the physically appropriate ingoing boundary condition for wave propagation. Because the accretion process considered here asymptotically settles to a stationary Schwarzschild spacetime of mass $m_2$, the final spacetime necessarily possesses a well-defined event horizon. Owing to its global and teleological nature, however, the location of the event horizon at any intermediate time cannot be inferred solely from local geometric quantities. Instead, its evolution must be determined by tracing the outgoing null generators backward in time from the final stationary configuration. Accordingly, the event-horizon trajectory satisfies the outgoing radial null geodesic equation
\begin{equation}\label{eq:horizonevol}
    \frac{\mathrm{d}r_H}{\mathrm{d}v}
    =
    \frac{1}{2}\left(1-\frac{2m(v)}{r_H}\right)~,
\end{equation}
which is integrated backward in time using the future boundary condition
\[
r_H(v_f)=2m_2,
\]
corresponding to the final Schwarzschild spacetime, where the event and trapping horizons coincide.

\begin{figure}[h]
    \centering
    \includegraphics[width=0.92\columnwidth]{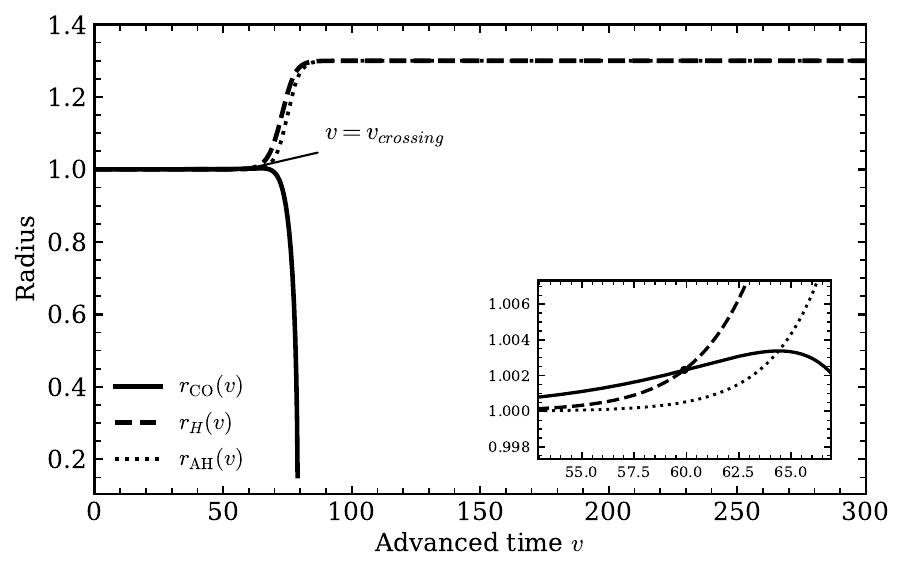}
    \caption{Evolution of the compact-object surface $r_{ECO}(v)$ (solid curve), the event horizon $r_H(v)$ (dashed curve), and the trapping horizon $r_{AH}(v)=2m(v)$ (dotted curve) during the accretion process. The inset shows a magnified view of the crossing region.}
    \label{fig:surface_horizon_evolution}
\end{figure}

The resulting evolution of the compact object surface and the event horizon is shown in Fig.~\ref{fig:surface_horizon_evolution}. Initially, the timelike surface lies outside both the trapping and event horizons, corresponding to a horizonless compact object. During the accretion process, the event horizon expands outward while the compact object surface evolves causally according to Eq.~(\ref{eq:ECOsurfaceevol}). The crossing of these two trajectories marks the onset of black hole formation, after which the physical surface lies inside the event horizon and is no longer causally connected to future null infinity. At sufficiently late times, both the event horizon and the trapping horizon asymptotically approach the Schwarzschild radius of the final stationary black hole.

With the background geometry, the timelike surface, and the event horizon trajectory completely specified, the dynamical spacetime is fully determined throughout the evolution. The quasinormal response is then obtained by numerically evolving scalar perturbations on this time-dependent background. This requires constructing the geometry on a double-null grid, evaluating the corresponding effective potential, and subsequently integrating the perturbation equation using the characteristic evolution scheme reviewed in the previous section. The numerical implementation employed to carry out these steps is described below.

\section{Numerical Implementation}\label{Sec:Num_Impl}

\begin{figure*}
\centering
\includegraphics[width=0.8\textwidth]{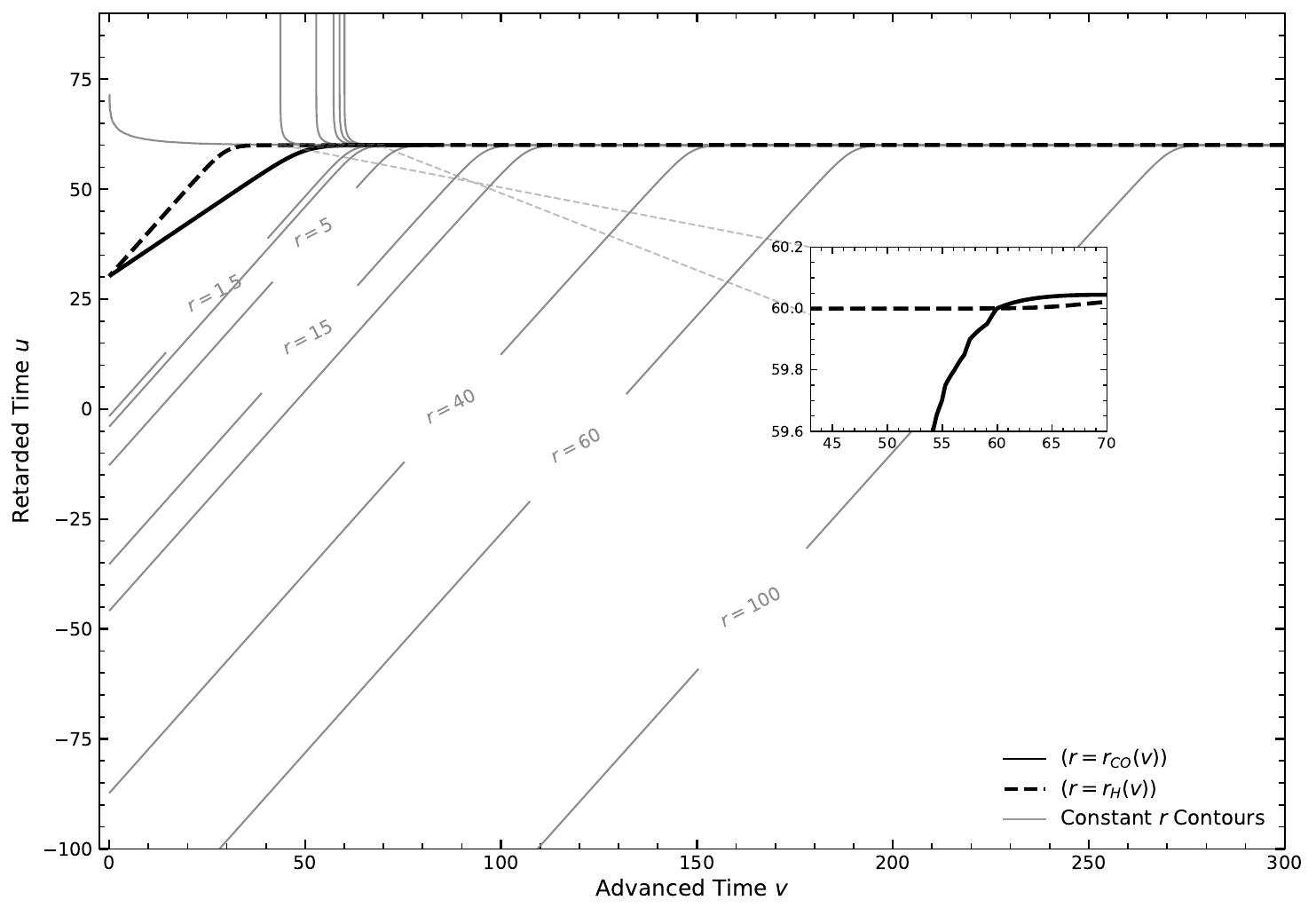}
\caption{Double-null representation of the dynamical spacetime showing constant-radius contours together with the trajectories of the ECO surface $r=r_{\rm ECO}(v)$ (solid) and the event horizon $r=r_H(v)$ (dashed). The event horizon separates outgoing null trajectories that eventually escape to infinity from those that are inevitably captured by the black hole region.}
\label{fig:uv_structure}
\end{figure*}

The numerical evolution proceeds in three successive stages. First, the dynamical background geometry is constructed from the prescribed mass profile and the evolution equations introduced in the previous section. The radial coordinate $r(u,v)$ is obtained by integrating Eq.~(\ref{eq:for_r}) along each outgoing null slice using the Eq. (\ref{eq:schwarztortoisecoord}) as the initial data on the reference hypersurface $v=v_0$. The corresponding metric function $f(u,v)$ is then computed from Eq.~\eqref{eq:for_f}, thereby completely specifying the Vaidya geometry in double-null coordinates. The timelike trajectory of the ECO surface and the event horizon trajectory are evolved independently using Eqs. (\ref{eq:ECOsurfaceevol}) and (\ref{eq:horizonevol}), respectively. The resulting double-null structure is illustrated in Fig.~\ref{fig:uv_structure}, where several constant-radius contours obtained from the numerical solution for $r(u,v)$ are shown together with the trajectories of the ECO surface and the event horizon. The instant at which the timelike surface intersects the event horizon is identified numerically and marks the transition from the horizonless compact object phase to the black hole phase.

Then, the effective potential is evaluated throughout the computational domain. At each point on the double-null grid, the potential is constructed from the instantaneous values of $m(v)$, $r(u,v)$, and $f(u,v)$. Prior to horizon formation, the interior of the compact object is excised from the computational domain by imposing a reflective boundary at the timelike surface. Following the formation of the black hole, this excision boundary is replaced by the event horizon, allowing the perturbation to propagate naturally into the black hole interior. Consequently, the computational domain evolves continuously from one appropriate for a horizonless compact object to one describing a black hole spacetime without introducing any discontinuity in the wave evolution.

With the dynamical potential specified, the scalar wave equation is evolved on the double-null grid using the second-order characteristic integration scheme \footnote{ The factor of $1/2$ in  Eq.~\eqref{eq_evolution} arises from the convention $ds^2=-2f(u,v)\,du\,dv+r^2d\Omega^2$ as opposed to the alternative convention $ds^2=-f(u,v)\,du\,dv+r^2d\Omega^2$, commonly used in the literature. The evolution equation in the other convention carries a factor of $1/8$ instead, see for example \cite{Chakravarti:2021clm}.},
\begin{equation}\label{eq_evolution}
\begin{aligned}
\psi(u,v)=&\,\psi(u-\Delta u,v)+\psi(u,v-\Delta v)-\psi(u-\Delta u,v-\Delta v)\\
&-\frac{\Delta u\Delta v}{2}
\Big[\psi(u-\Delta u,v)V(u-\Delta u,v)\\
&\hspace{3.1cm}+\psi(u,v-\Delta v)V(u,v-\Delta v)
\Big].
\end{aligned}
\end{equation}
Initial data are prescribed on two intersecting null hypersurfaces, namely an outgoing surface $u=u_0$ and an ingoing surface $v=v_0$. On the former, we specify a sinusoidally modulated Gaussian pulse,
\begin{equation}
    \psi(u=u_0,v) = \exp\left[-\frac{(v-v_c)^2}{2\sigma^2}\right]\sin(\omega v),
\end{equation}
with $v_c = 10$, $\sigma=3$, and $\omega=0.25$, while the perturbation is taken to vanish on the latter, $\psi(u,v=v_0) = 0$. The oscillatory modulation is introduced primarily for numerical convenience, as it facilitates the excitation and identification of the dominant quasinormal modes in the subsequent evolution. The scalar field is subsequently extracted at a fixed areal radius sufficiently outside the compact object. Since the observer remains at a constant physical radius throughout the evolution, the extracted signal directly captures the continuous transition of the quasinormal response from the initial horizonless configuration to the final black hole state.

\section{Results}\label{Sec:Res}

In this section, we present the numerical results obtained from the characteristic evolution of scalar perturbations on both black hole and exotic compact object backgrounds. We begin by considering the black hole case, for which the quasinormal response of both stationary Schwarzschild and accreting Vaidya spacetimes has been extensively studied in the literature. The purpose of this comparison is twofold. First, it serves as a validation of our numerical implementation by reproducing the qualitative features previously reported for dynamical black hole ringdown. Second, it provides a useful baseline against which the effects associated with horizon formation from an initially horizonless compact object can be compared.

\begin{figure*}
\centering
\includegraphics[width=0.7\textwidth]{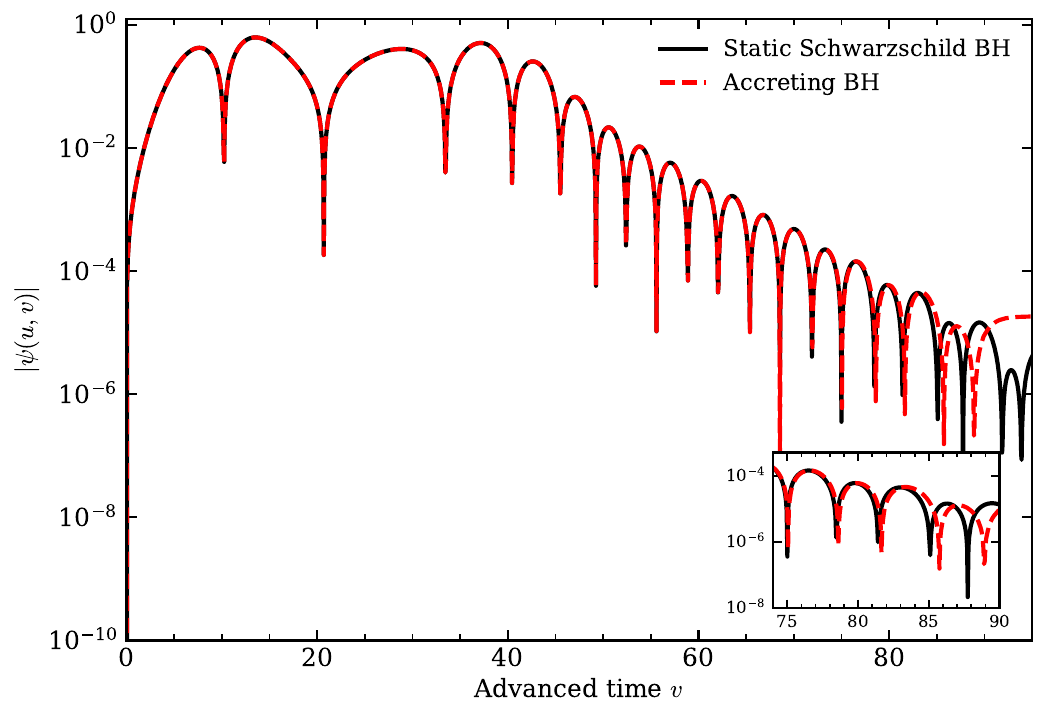}
\caption{
Comparison of scalar field waveforms extracted at a fixed observer radius $r_{\rm obs}=11.0$ for a static Schwarzschild black hole and an accreting Vaidya black hole. The two signals remain nearly identical during the early stages of evolution, while significant accretion leads to a gradual departure from the static waveform. The accreting configuration exhibits weaker damping and a small phase shift relative to the static case, as highlighted in the inset.
}
\label{fig:waveforms}
\end{figure*}

Fig.~\ref{fig:waveforms} compares the scalar waveforms for a static Schwarzschild black hole with those of an accreting Vaidya black hole. The static Schwarzschild black hole exhibits the familiar quasinormal response characterized by exponentially damped oscillations with an approximately constant oscillation period and decay rate. Since the background spacetime is stationary, the waveform is entirely determined by the fixed black hole mass $m_0 = 0.5$. The accreting Vaidya black hole displays a qualitatively similar ringdown signal during the early stages of evolution, with the two waveforms remaining nearly indistinguishable before the accretion becomes appreciable. As the accretion enters its rapid growth phase near $v=v_1$, the waveform gradually departs from the static Schwarzschild signal. In particular, the accreting configuration exhibits a systematically slower decay and develops a small but noticeable phase shift relative to the static case. These features are qualitatively consistent with previous investigations of quasinormal ringing in Vaidya spacetimes, which showed that an increase in black hole mass generally leads to weaker damping and modifies the quasinormal frequencies of the system \cite{Abdalla:2006vb, Capuano:2026tjy, Capuano:2024qhv}. The observed phase shift further reflects the nonstationary nature of the background geometry and the finite response time of the perturbation to the evolving scattering potential. The waveform is extracted at a fixed observer radius $r_{\rm obs}=11.0$, corresponding to $r_{\rm obs}/m_0 = 22$ initially and $r_{\rm obs}/m_f \simeq 16.9$ in the final state, ensuring that the observer remains sufficiently far outside both the horizon and the photon sphere throughout the evolution. Consequently, the extracted signal provides a faithful representation of the outgoing quasinormal response of the dynamical spacetime.

\begin{figure*}
\centering
\includegraphics[width=0.9\textwidth]{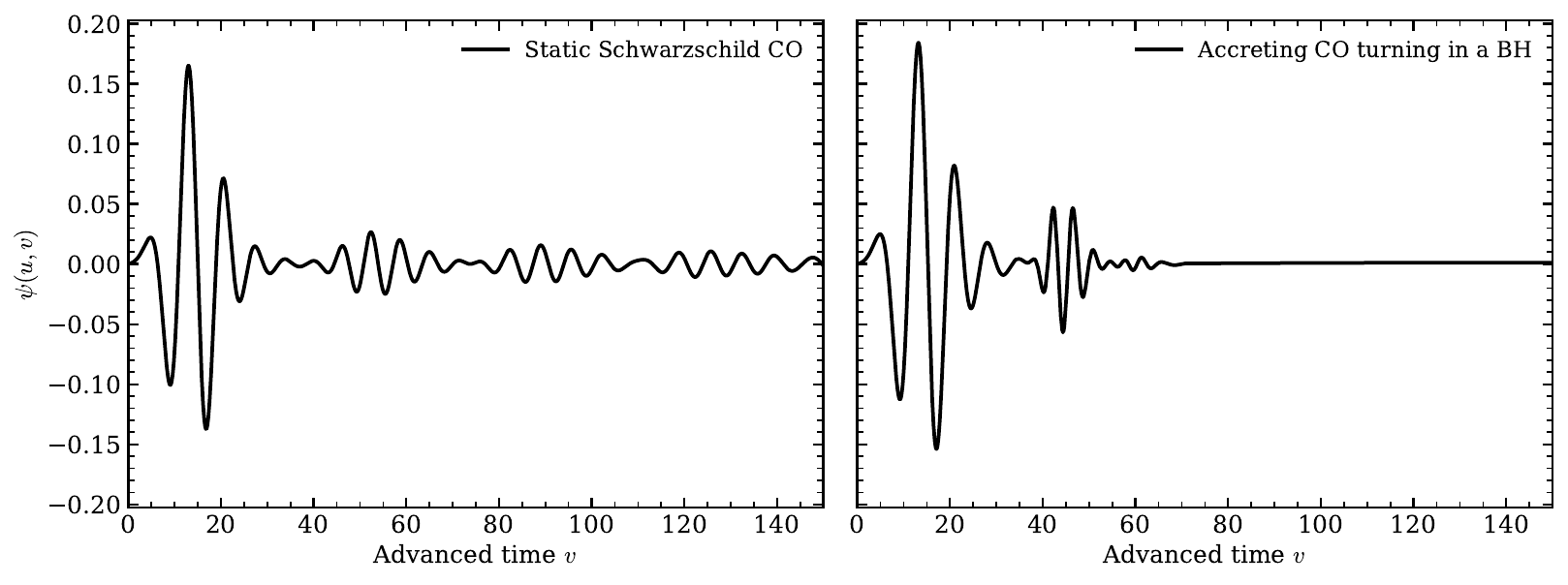}
\caption{Scalar field waveforms extracted at a fixed observer radius $r=r_{\rm obs}=2$. Left panel: Signal for a static Schwarzschild compact object possessing a reflecting surface, exhibiting persistent late-time echoes. Right panel: Corresponding signal for an accreting compact object that eventually undergoes a transition to a black hole. The early-time waveform displays echoes characteristic of a horizonless configuration, while the subsequent formation of an event horizon suppresses further reflections and the signal gradually approaches the conventional black hole ringdown. The transition from an echo-dominated signal to a standard ringdown therefore provides a characteristic signature of the dynamical emergence of an event horizon.
}
\label{fig:ecotoBH}
\end{figure*}
We now consider the corresponding evolution for compact objects. In contrast to the black hole case discussed above, the compact object initially possesses a timelike reflecting surface and only later undergoes horizon formation. This setup allows us to directly investigate how the emergence of an event horizon modifies the late-time waveform. Fig.~\ref{fig:ecotoBH} compares the scalar waveforms of a static ECO and a Vaidya ECO that eventually collapses into a black hole. The waveform is extracted at a fixed observer radius $r_{\rm obs}=2.0$, which remains outside the compact object surface and the event horizon throughout the evolution while being sufficiently close to the effective scattering region to allow the late-time echo signal to be resolved accurately. At substantially larger extraction radii, the amplitudes of successive echoes become increasingly suppressed, rendering the late-time echo structure difficult to resolve numerically. The left panel corresponds to a static compact object possessing a perfectly reflecting surface located at $$r_{ECO} = 2m_0\left(1+\epsilon\right), \:\: ~\epsilon = 10^{-7}.$$. The waveform exhibits a sequence of late-time echoes. These echoes arise because part of the scalar radiation becomes repeatedly trapped between the effective photon-sphere potential barrier and the reflecting compact object surface. Each successive reflection generates an additional pulse in the observed signal, as shown in the left panel of Fig.~\ref{fig:ecotoBH}, producing an echo train whose amplitude gradually decreases with time owing to the partial escape of radiation at every scattering event.

The right panel depicts the corresponding evolution for an accreting compact object whose surface eventually intersects the event horizon. Prior to the formation of the horizon, outgoing perturbations can repeatedly reflect from the ECO surface and return to the observer. Consequently, the early time waveform closely resembles that of the static compact object and displays clear echo signatures. This indicates that, before horizon formation, the spacetime behaves observationally as a horizonless ECO. A notable feature visible in the waveform is the progressive compression and suppression of the echo sequence during the accretion phase. Compared to the static compact object case, successive echoes become increasingly closely spaced and less pronounced as the system approaches horizon formation. These effects reflect the dynamical modification of the effective cavity between the compact object surface and the photon-sphere potential barrier.

As accretion proceeds, the compact object surface approaches the growing event horizon and eventually becomes enclosed by the growing event horizon, as illustrated in Fig. \ref{fig:uv_structure}. Once this occurs, the reflecting boundary ceases to communicate causally with the exterior spacetime, and radiation that would previously have been reflected back toward the observer is instead absorbed by the newly formed horizon. Consequently, the echo sequence is terminated, and the waveform undergoes a smooth transition to the conventional exponentially damped black hole ringdown. This transition may become a potentially observable signature of black hole formation in accreting ECOs and may serve as a useful probe of the dynamical emergence of black holes. We also remark that the ringing phase of binary mergers in dirty environments constitutes the closest physical scenario that matches our constructed framework \cite{Leung:1997was, Cardoso:2021wlq, Spieksma:2024voy, Berti:2018vdi}.

\section{Conclusion and Discussion}\label{Sec:Disc}

In this work, we have investigated the evolution of the quasinormal mode response of an initially horizonless compact object undergoing continuous accretion with the exterior geometry modeled as an ingoing Vaidya spacetime. By combining the causal horizon-formation scenario developed in Ref.~\cite{Carballo-Rubio:2018vin} with the double-null numerical framework for perturbations in dynamical Vaidya geometries, we have provided, to our knowledge, the first dynamical study of how gravitational-wave echoes evolve as an ECO is transformed into a black hole. We capture the distinct signature of the ECO lifecycle, specifically the initial emergence of GW-echoes and their eventual disappearance, earmarking a key novelty of our work.

Our analysis shows that sustained accretion inevitably drives sufficiently compact horizonless objects toward black hole formation. As the mass of the object increases, the event horizon expands and eventually overtakes the timelike surface of the compact object. Once this occurs, the reflecting surface responsible for producing late-time echoes becomes causally disconnected from future null infinity. Consequently, the echo cavity, formed between the photon-sphere potential barrier and the physical surface, disappears dynamically. Rather than persisting indefinitely, the echo train terminates naturally as the event horizon engulfs the surface, after which the waveform smoothly approaches the standard quasinormal ringdown expected for an ordinary black hole.

This behavior provides a characteristic observational signature of the horizon-formation process itself. Instead of treating black holes and ECOs as two distinct classes of compact objects, our results demonstrate that there may exist an intermediate dynamical phase during which an initially horizonless object exhibits echoes before continuously evolving into a conventional black hole. Such a transition produces a time-dependent modification of the ringdown signal that cannot be captured by analyses based solely on stationary background geometries. Future gravitational-wave observations with increased post-merger sensitivity may therefore be capable not only of searching for echoes, but also of constraining the dynamical disappearance of echoes resulting from accretion-driven horizon formation. In fact, because mergers involving accreting supermassive black holes are expected to be observed with a high signal-to-noise ratio by LISA, the dynamical transition from an echo-dominated waveform to a conventional black hole ringdown identified here may provide an interesting observational signature of horizon formation.

Our results also reinforce the causal argument of Ref.~\cite{Carballo-Rubio:2018vin}. Even if horizonless compact objects exist in nature, sufficiently compact configurations cannot remain horizonless under realistic accretion for extended periods. The present work demonstrates that this causal instability may have an equally important dynamical consequence in the gravitational-wave sector: the observable quasinormal response changes qualitatively once the event horizon overtakes the compact object surface. In this sense, the disappearance of echoes can be viewed as an observational manifestation of the causal transition from an ECO to a black hole.

At this point a word about the detectability of the `disappearing echo' spectrum is in order. The network SNR of GWs depend upon among other things, the remnant mass as well as the detector sensitivity. A GW150914-like event has a standalone ringdown network SNR of $\sim8.5$ \cite{Tsang:2019zra}, while the same number for a GW250114-like event is $\sim18$ \cite{Akyuz:2025seg} or nearly twice as large. From Fig.~\ref{fig:ecotoBH} we observe that the echo features have a time-domain amplitude which is smaller than the primary QNM oscillatory feature by a factor of $\sim4-5$. This means that even GW250114-like events will produce a standalone echo SNR of just $\mathcal{O}(1)$, while those from GW150914-like events cannot be distinguished from detector noise. Therefore we cannot expect a distinguishable signature of ECO-BH transition for stellar mass mergers with current detector sensitivity. Future ground-based detector networks like the Einstein Telescope or Cosmic Explorer stand a much better chance of resolving the GW-echoes. Additionally detectors like LISA, TianQin or DECIGO can also be useful to check for accreting ECO signatures with EMRIs.

Several extensions of the present work would be of considerable interest. The analysis presented here has been restricted to scalar perturbations propagating on a spherically symmetric Vaidya background. A natural next step is to investigate gravitational and electromagnetic perturbations, whose spectra are directly relevant for astrophysical observations. It would also be worthwhile to extend the analysis to rotating dynamical spacetimes, where frame dragging, superradiance, and the richer structure of Kerr quasinormal modes may significantly modify the transition. Furthermore, incorporating more realistic accretion histories and matter distributions, together with waveform modeling suitable for matched-filter searches, would allow the present framework to be connected directly with future gravitational-wave observations.

Overall, our work establishes a dynamical framework for studying quasinormal modes and gravitational-wave echoes during the evolution of a dynamical ECO and the formation of the event horizon. The results suggest that the appearance and disappearance of echoes is itself a robust and potentially observable signature of black hole formation. We therefore expect that future high-precision gravitational-wave detectors, including next-generation ground-based observatories and space-based missions such as LISA, may provide a unique opportunity to probe not only the existence of exotic compact objects but also their possible dynamical conversion into astrophysical black holes.

\section*{Acknowledgment} The authors thank C. Chirenti and A. Saa for clarification on an important aspect of the accretion process necessary to compute the QNMs/GW-echoes. C.S. acknowledges hospitality and local support from CMI, Chennai, and IACS, Kolkata. The research of K.C is funded by the Ramanujan Fellowship (RJF/2025/000175) scheme of ANRF, Govt. of India. K.C. acknowledges hospitality and local support from IACS, Kolkata, and IIT Gandhinagar. S.S.’s research is supported by the Department of Science and Technology, Government of India, under the ANRF CRG Grant (No. CRG/2023/000934).
\bibliographystyle{apsrev4-1}
\bibliography{Vaidya_ECO}

\end{document}